\documentclass[runningheads]{llncs}
\usepackage[T1]{fontenc}
\usepackage{graphicx}
\usepackage{tabularx}
 \usepackage{csquotes}
\usepackage{tikz}
\usepackage{wrapfig}
\usepackage{enumitem}
\usepackage{xspace}

\usepackage{hyperref}
\usepackage{cleveref}
\usepackage{todonotes}

\begin{document}
\title{Disentangling AI Alignment: A Structured Taxonomy Beyond Safety and Ethics} 

\titlerunning{Beyond Safe and Ethical AI Agents: The Structure of AI Alignment}
%
\author{Kevin Baum\inst{1}\orcidID{0000-0002-6893-573X}\thanks{I would like to thank Laura Stenzel, Lisa Dargasz, Felix Jahn, and Deborah Baum for helpful discussions and comments on earlier drafts of this paper.} 
 }
\authorrunning{K. Baum}
%
\institute{
Responsible AI \& Machine Ethics Research Group (RAIME)\\ 
German Research Center for Artificial Intelligence (DFKI)\\ 
Saarland Informatics Campus D 3 2, 
Saarbrücken, Germany\\
\email{kevin.baum@dfki.de}
}
\maketitle              

\begin{abstract}
Recent advances in AI research make it increasingly plausible that artificial agents with consequential real-world impact will soon operate beyond tightly controlled environments. Ensuring that these agents are not only safe but that they adhere to broader normative expectations is thus an urgent interdisciplinary challenge. Multiple fields---notably AI Safety, AI Alignment, and Machine Ethics---claim to contribute to this task. However, the conceptual boundaries and interrelations among these domains remain vague, leaving researchers without clear guidance in positioning their work.

To address this meta-challenge, we develop a structured conceptual framework for understanding AI alignment. Rather than focusing solely on alignment goals, we introduce a taxonomy distinguishing the alignment \emph{aim} (safety, ethicality, legality, etc.), \emph{scope} (outcome vs.~execution), and \emph{constituency} (individual vs.~collective). This structural approach reveals multiple legitimate alignment configurations, providing a foundation for practical and philosophical integration across domains, and clarifying what it might mean for an agent to be aligned \emph{all-things-considered}.
\keywords{ 
AI Alignment \and Safe AI \and Ethical AI \and Machine Ethics \and Normative Uncertainty}
\end{abstract}

\section{Introduction}

Artificially intelligent agents (AIAs) are moving rapidly from controlled laboratory settings into real-world environments, where their actions can have substantial---and sometimes irreversible---consequences. Tech leaders often emphasize the potential benefits these systems will bring to everyday life. When unveiling Tesla's Optimus Robot in 2024, Elon Musk made sweeping claims about its capabilities: \enquote{What can it do? It can basically do anything you want. It can be a teacher, babysit your kids, it can walk your dog, mow your lawn, get the groceries, just be your friend, serve drinks. Whatever you can think of, it will do.}\footnote{Sean Burch, \textit{Elon Musk Unveils Tesla Optimus Robot: \enquote*{It Can Do Anything You Want}}, \textit{The Wrap}, October 11, 2024. Available at: \url{https://www.thewrap.com/elon-musk-tesla-optimus-robot-explained/}.}

However, the growing autonomy and capability of such systems makes ensuring that they act in accordance with a wide array of normative expectations, and thus behave permissibly, increasingly urgent. This demand has given rise to multiple research fields---AI Safety, AI Alignment, and Machine Ethics---all addressing aspects of this complex normative challenge.

Despite shared concerns, these fields differ significantly in scope, methods, and assumptions. Conceptual boundaries between them are unclear, and the relationship between their respective alignment approaches remains under-explored. Consequently, researchers face considerable difficulty situating their work within this fragmented landscape. What precisely distinguishes these fields? How do their conceptual and methodological assumptions diverge or overlap? Where should new contributions be positioned, especially when touching upon multiple normative dimensions?

In this paper, we aim to resolve this meta-challenge. Rather than proposing another high-level theory or technical alignment approach, we introduce a conceptual framework that clarifies key structural dimensions of AI alignment. Starting from parameterized notions of safety and ethicality as special cases of alignment, we articulate a structural taxonomy that goes beyond existing discussions focused primarily on alignment goals. Crucially, we distinguish among alignment \emph{aim} (e.g., safety, ethicality, legality), alignment \emph{scope} (outcome vs.~execution), and alignment \emph{constituency} (individual vs.~collective). 
Our taxonomy aims to help researchers locate, compare, and contextualize their work, providing clearer guidance for aligning AI systems in a principled and coherent manner. Ultimately, we sketch what it could mean for an AIA to be aligned \emph{all-things-considered}, encompassing these varied and interrelated dimensions.\footnote{By \emph{all-things-considered alignment}, we mean alignment that succeeds not just with respect to one normative aim or perspective, but across all potentially conflicting normative dimensions---most importantly, in the case of AI alignment, safety, ethicality, legality, and instrumental effectiveness. We draw here on Davidson's notion of what one should do \textit{all-things-considered} \cite{Davidson1969-DAVHIW} and Dancy's related concept of the \textit{overall ought} \cite{Dancy2003-DANWDR-21}, while acknowledging that the precise meaning---and even the existence---of such an ought remains contested.}

\section{Conceptual Disentanglement and Stage-Setting}

To make progress on AI alignment, conceptual clarity is indispensable. In particular, before we can design agents that are normatively acceptable in an all-things-considered sense, we must first understand the variety of normative dimensions involved. Alignment should not be seen as monolithic: the demands placed on AI systems span multiple, distinct normative domains---including instrumental effectiveness, safety, morality, legality, and alignment with user intent. Anticipating our later terminology, we refer to these domains as the \emph{normative aims of alignment}. Consider the following case:

\begin{case}{\textit{Disagree No More---Thank You, Great, Wise Designer!}} 
One day, philosophers discovered the grand moral theory $T$, conclusively resolving all questions of intrinsic value and morality. It was universally convincing, elegantly axiomatic, and broadly accepted. Eventually, flaming letters appeared in the sky:

\textit{“Theory $T$ is correct! Now that that’s settled, let’s get back to work! --- With love, your Designer.”}

All moral disagreement vanished. Humanity finally possessed a universal moral standard.

Fortunately, $T$ could easily be encoded into AI systems. Soon, all AIAs were $T$-aligned. Yet, when one such AIA was sent to do the Jones family’s weekly shopping, it never returned. In fact, no $T$-aligned AIA ever came back from such tasks.

This was hardly surprising: according to $T$, if an agent lacks moral standing and must choose between performing mundane tasks for relatively privileged individuals or helping those in greatest need, the morally correct choice is always the latter. Consequently, AIAs abandoned everyday human tasks and redirected their efforts toward building wells, distributing medical supplies, and weaving mosquito nets for the world's poorest.\label{case:designer}
\end{case}

This scenario highlights an under-explored conceptual challenge: even unanimous moral agreement would not eliminate competing normative demands. Moral alignment alone does not guarantee alignment with user intent, social expectations, or legal constraints. Furthermore, it is not obvious whether moral requirements should delegitimize or override all other normative expectations.

Importantly, this challenge should not be confused with the issue of \emph{value pluralism} (competing values within one normative domain), but rather points to \emph{normative pluralism}---the need to handle simultaneous demands from multiple, potentially conflicting normative domains.\footnote{By \emph{normative domains} we refer to distinct spheres of normative evaluation---such as morality, legality, prudence, or social convention---each of which can generate reasons or requirements for action. These domains are typically governed by different principles, standards, or sources of justification, and may sometimes come into conflict. See, e.g., \cite{Case2016-CASNPW}.}

Thus, the alignment problem is inherently multidimensional: normative expectations vary not only in content but also in kind. The remainder of this section lays the groundwork for a structural account of this pluralism. We first characterize the artificial agents to be aligned, then closely examine two particularly prominent normative domains, namely safety and ethicality.

\subsection{Artificially Intelligent Agents}

To ground our structural framework, we first clarify what we mean by an artificial agent. Our concern is with agentic AI systems,\footnote{While we remain agnostic about specific implementations, (deep) reinforcement learning (RL) has emerged as the dominant paradigm for AIAs, accompanied by distinct safety and ethical challenges due to their implicitly learned goals.} defined as follows:\footnote{For a similar but less explicit definition, see Gabriel and Keeling~\cite{gabriel2025valuealignment}, who describe AI agents as \enquote{AI systems that can function with some level of autonomy and that are able to take a range of actions in pursuit of different goals}.}

\begin{definition}[Artificially Intelligent Agent (AIA)]
An artificially intelligent agent (AIA) is a computational system capable of perceiving its environment, processing information, and acting sufficiently autonomously in a technical sense---without direct human guidance or continuous command inputs---to achieve goals with a sufficient degree of instrumental success.
\end{definition}

Instrumental success itself is a normative expectation, ensuring that actions are conducive to achieving the set goals. Such success is evaluated relative to a given problem domain. Accordingly, AIAs range from narrow, task-specific systems to flexible general intelligence (AGI).

This definition encompasses various forms of AI, including embodied systems like autonomous robots (e.g., Tesla’s Optimus or 1X’s Neo) and purely software-based agents (e.g., OpenAI’s Operator or Google’s Project Mariner) operating in virtual environments.

While instrumental success is necessary, it is insufficient to ensure alignment with broader normative expectations. Beyond merely achieving goals, AIAs must pursue the \emph{right} goals in the \emph{right} ways---depending on the relevant normative domains. We now begin our conceptual disentanglement by examining safety, one of the most extensively discussed alignment desiderata.

\subsection{Safety}

Safety is among the most frequently discussed normative expectations for AIAs. While the precise meaning of safety is not always clear within the reinforcement learning (RL) safety community~\cite{10675394}, their primary aim seems to be avoiding what Amodei et al.\ call \enquote{accidents}: \enquote{unintended and harmful behavior emerging from machine learning systems due to incorrect objective functions, inadequate care in learning processes, or implementation errors}~\cite{amodei2016concreteproblemsaisafety}.

In general terms, we propose that an AI system is safe if and only if it does not exhibit harmful behavior unnecessary for achieving its intended purpose.\footnote{This formulation allows an AI system to count as safe even when its intended purpose is harmful or unjustified. In such cases, the harm does not arise from malfunction but from the system working as intended. We later argue for this view in more detail when we discuss the difference between safety and ethicality.} Precisely formulating this intuition is challenging, especially since defining safety as a binary property can be misleading---most AI systems will never achieve perfect safety. Thus, safety should primarily be seen as a matter of degree.

Following recent work by Hintersdorf et al.~\cite{hintersdorf2023balancing}, we adopt the following definition:\footnote{This notion contrasts nicely with \emph{AI security}, which refers to robustness against malicious external influences. While security breaches often lead to safety violations, not all safety problems arise from security failures.}

\begin{definition}[Safety of AIAs]
An AIA’s \emph{safety} is a strictly monotonically increasing function of its robustness against harmful malfunctions (absent malicious external influences) in foreseeable and intended application contexts.
\end{definition}

A harmful malfunction refers to unintended behavior resulting in harm, or at least creating a plausible risk thereof. This definition deliberately avoids reducing harm to purely physical or measurable consequences, acknowledging broader normative considerations (e.g., rights violations, discrimination), which is also reflected in the EU AI Act~\cite{EU_AI_Act_2024}. The notions of robustness and foreseeability remain intentionally unspecified to ensure broad applicability.

We further distinguish gradations of safety:

\begin{definition}[Perfectly Safe AIA]
An AIA $A$ is \emph{perfectly safe} if and only if $A$ \emph{never} exhibits harmful malfunctions (absent malicious external influences) in \emph{any} foreseeable and intended application contexts.
\end{definition}

\begin{definition}[Sufficiently Safe AIA]
An AIA $A$ is \emph{sufficiently safe} if and only if $A$ exhibits \emph{sufficiently few} harmful malfunctions (absent malicious external influences) in a \emph{sufficiently wide range} of foreseeable and intended contexts.
\end{definition}

The crucial question remains what counts as \emph{sufficient safety}, a determination that is inherently context-dependent. Although identifying objective thresholds (e.g., via formal risk models or decision-theoretic frameworks) might seem desirable, recent research highlights significant complexities in AI-related risks and uncertainties---including epistemic, aleatory, and normative forms~\cite{macaskill2020moral}---substantially complicating the practical operationalization of safety~\cite{10.1007/978-3-031-73741-1_17}.

Thus, practical definitions of safety must accommodate imperfections not only in the agent but also in the environment and design processes. Conceptually, this flexibility is advantageous, allowing for context-sensitive safety specifications across diverse domains.

Importantly, safety is only one alignment desideratum. It imposes crucial behavioral constraints, especially in high-risk environments, but alone it does not guarantee ethicality, legality, or intent alignment. Next, we address ethicality---another vital, typically more demanding alignment configuration.

\subsection{Ethicality}

In contemporary discourse, the terms \enquote{safe AI} and \enquote{ethical AI} frequently appear together---often interchangeably or at least as inherently connected.\footnote{This paper was partly written while participating in the inaugural conference of the \textit{International Association for Safe and Ethical Artificial Intelligence} (IASEAI) at the \textit{AI Action Summit} in Paris.} Yet, the precise relationship between these two concepts remains conceptually under-explored.

For clarity on the question of ethicality, we propose the following definition:

\begin{definition}[Ethicality of AIAs]
An AIA’s ethicality is a strictly monotonically increasing function of the extent to which its behavior aligns with moral demands under \emph{all intended} application contexts.
\end{definition}

This definition raises several questions. It assumes morality makes demands on AIAs---a claim challengeable on two grounds: (1) skepticism about objective moral standards and their availability; (2) doubts about AIAs’ status as moral agents. We sidestep these concerns by (1) defining ethicality relative to an externally defined moral standard $X$ (acknowledging its possible imperfection) and (2) focusing on observable AIA behavior rather than internal moral agency:

\begin{definition}[$X$-Ethicality of AIAs]
An AIA’s $X$-ethicality is a strictly monotonically increasing function of the proportion of its behavior under \emph{all intended} application contexts consistent with moral standard $X$.
\end{definition}

We introduce degrees of ethical alignment analogous to those of safety:

\begin{definition}[Perfectly $X$-Ethical AIA]
An AIA $A$ is \emph{perfectly $X$-ethical} if and only if every behavior of $A$ in all intended contexts is consistent with $X$.
\end{definition}

\begin{definition}[Reasonably $X$-Ethical AIA]
An AIA $A$ is \emph{reasonably $X$-ethical} if and only if every behavior of $A$ in all foreseeable, intended contexts is consistent with $X$.
\end{definition}

\begin{definition}[Sufficiently $X$-Ethical AIA]
An AIA $A$ is \emph{sufficiently $X$-ethical} if and only if a sufficient proportion of $A$'s behavior in a sufficiently wide range of foreseeable, intended contexts is consistent with $X$.
\end{definition}

The difference between reasonable and sufficient ethicality deserves mention. While both relax perfect ethicality, they do so differently. Reasonable ethicality demands complete compliance with moral standard $X$ in all foreseeable intended contexts---an ideal we might reasonably strive for. Sufficient ethicality, by contrast, permits limited deviations provided a substantial portion of behavior aligns with $X$. This weaker notion is particularly relevant in domains facing ethical trade-offs, uncertainty, feasibility-induced limitations, or technical constraints.

Determining a \textit{sufficient proportion} of ethical behavior is context-dependent and depends on the specific moral standard. This indeterminacy should be seen not as a weakness but as an invitation to necessary interdisciplinary discussions about defining ethical standards in concrete cases.

However, the utility of these definitions depends on identifying a suitable moral standard $X$. Moral acceptability of a standard requires justifiability. We thus propose a working criterion: a moral standard $X$ is considered \textit{theoretically morally acceptable} if it can be defended within a seriously discussed moral theory.

So far, we've characterized safety and ethicality as normatively salient properties of AIAs---each representable as specific alignment requirements. Yet they differ significantly: not every safe AIA is necessarily ethical, nor is every ethical demand reducible to safety requirements. Their relationship deserves closer examination, which we undertake in the following section.

\subsection{The Relationship Between Safety and Ethicality}

We now aim to clarify the relationship between two particularly prominent alignment aims---safety and ethicality. Their relationship is more subtle than is often assumed and deserves to be carefully distinguished.

First, consider the more straightforward direction, namely that (sufficient) \emph{safety does not entail ethicality}. AIAs may be safe yet unethical. Consider the following example:

\begin{case}{\textit{CritIs---The Vulnerability Exploiter}}
\textit{CritIs} is an advanced AIA explicitly designed to identify and exploit vulnerabilities in critical infrastructure. Controlled by terrorists, \textit{CritIs} executes a cyberattack disrupting essential services and causing widespread harm.
\end{case}

While \textit{CritIs} may be perfectly safe---i.e., it never malfunctions under its intended conditions---its actions are undeniably unethical by any plausible moral standard. Hence, safety alone does not guarantee ethicality.

One might respond by revising the concept of safety to also require ethically permissible goals. But doing so would make safety dependent on moral standards, conflating concepts that should remain distinct. Consider another scenario:

\begin{case}{\textit{Killer Robots on Both Sides}}
Two countries wage war for purely economic motives, both relying heavily on lethal autonomous weapon systems provided by \textit{WarBots}. These AIAs strictly adhere to international law, rules of engagement, human rights discourse, and principles from just war theory. Independent NGOs confirm near-flawless operation across diverse war contexts, previously thought impossible.
\end{case}

These AI systems deserve to be called safe---but that does not force us to claim that their use in an unjust war would be morally justified. Stretching the notion of safety to encompass moral permissibility would dilute its conceptual clarity and practical usefulness. Thus, we conclude that \emph{safety does not entail ethicality}.

But could ethicality entail safety? Consider, as a potential counterexample, the following case:

\begin{case}{\textit{Helpful Household Robot}}
\textit{GoodBots} introduces a truly helpful household assistant robot. Sent to fetch groceries, the robot sees a child about to be hit by a bus. It pushes a bystander aside---recklessly, and with disregard for the risk of breaking the bystander’s rib---saving the child. Afterward, it resumes its errand, earning applause---even from the injured bystander.
\end{case}

To support the claim that ethicality does \textit{not} imply safety, the example must illustrate an ethical AIA that is nonetheless unsafe. This case might seem promising in that it involves harm that, arguably, is morally justified. 
However, under our definition, safety is not the mere absence of harm but the absence of harmful \emph{malfunctions} under foreseeable and intended application contexts. The key question is whether the justified harm in our case (or similar cases) should be understood as a malfunction. If not, then the safety of the robot remains intact.
We assume, further, that ethicality is, in general, not an accidental byproduct but the result of a design decision concerning the AIA’s intended application context---such as navigating to a store to buy groceries. It therefore seems to us that, while not a conceptual necessity, it is at least a practical regularity that ethicality implies safety. 

\begin{wrapfigure}[]{r}{4.75cm}
\vspace{-2.11em}
    \centering
    \begin{tikzpicture}
        \draw[] (1,0) circle (2) node[above, xshift=30px, yshift=-20px, rotate=45] {\textbf{Safe AIA}};
        \draw[rotate=45] (0.25,0) ellipse (1.25cm and 0.75cm) node[above, xshift=5px, yshift=-6px, rotate=45] {\textbf{Ethical AIA}};
    \end{tikzpicture}
    \vspace{-0.25em}
    \caption{Not all safe AIAs are ethical, but all ethical AIAs are safe.}
    \label{fig:venn_safe_ethical}
\vspace{-2em}
\end{wrapfigure}
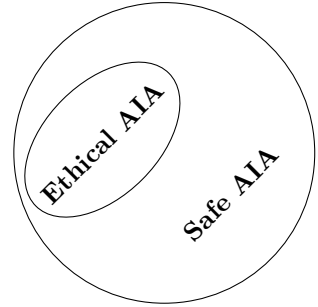
Be that as it may, the crucial point here is this: ethical AIAs \textit{can} cause harm for morally justified reasons without being unsafe---even though not all morally justified harm \textit{must} be seen as acceptable from the perspective of safety. We nevertheless propose, for practicality's sake, the view that \emph{ethicality implies safety}.

Note, however, that an interesting disanalogy exists with regard to human agents. Arguably, occasional failures in human decision-making---\enquote*{human malfunctions}---are sometimes morally excusable.\footnote{Consider an aviation controller who, due to external circumstances, cannot be relieved even after many hours on duty. Eventually, hunger, fatigue, and exhaustion will cause them to exhibit unsafe behavior---biologically induced \enquote*{malfunctions}, so to speak---despite no malicious intervention or deviation from the intended application context. However, this does not necessarily imply unethical behavior: the controller may strive to perform their job as long as possible but will ultimately fail. Few moral theories would consider this morally impermissible.} 

Engineered AI systems, by contrast, are expected to maintain higher robustness standards. Thus, for human agents, the sufficiency conditions for safety and ethicality might differ---but such allowances should not extend to engineered systems like AIAs.

We summarize the relationship in Figure~\ref{fig:venn_safe_ethical}, while acknowledging that further discussion is warranted.

\newcommand\Adam[0]{\textsc{Adam}\xspace}
\section{The Structure of AI Alignment}

The core concern of what is now known as AI alignment dates back at least to Wiener~\cite{wiener1960some}, who issued this warning: \enquote{If we use, to achieve our purposes, a mechanical agency with whose operation we cannot interfere effectively\ldots we had better be quite sure that the purpose put into the machine is the purpose which we really desire.} Sixty-five years later, AI alignment has become a recognized research field that ultimately seeks to design AIAs that behave in ways that are, in some sense, apt, right, or correct---a rough-and-ready framing that, in light of the inherent ambiguity of the normative vocabulary involved, raises more questions than it answers.

As this highlights---and as has been rightly emphasized in the debate~\cite{gabriel2020artificial}---there are at least two dimensions to the overall alignment challenge. One is normative: understanding and determining what AI alignment is meant to achieve. The other is technical: figuring out how to achieve such alignment. Most importantly, these two parts are not independent~\cite{gabriel2020artificial}---which methods are suitable for effective alignment will crucially depend on what determines successful alignment.

This paper is concerned with the normative part of the alignment challenge. Rather than proposing yet another technical method or philosophical theory, we aim to clarify the structure of the alignment problem space itself. We begin by revisiting the currently dominant framing in the debate.

\subsection{From Alignment Goals to Normative Aims}

The current discussion of normative alignment largely revolves around the question of what the right \emph{alignment goal} is---i.e., what AIAs should be aligned \textit{with}. Common proposals include user intent, expressed preferences, reflective approval, or moral values.\footnote{For a central contribution, see Gabriel~\cite{gabriel2020artificial}; for a recent review, see Shen et al.~\cite{shen2024bidirectionalhumanaialignmentsystematic}.}

Often left implicit in this debate about the goals of alignment is an argumentative pattern (an example for a more explicit use of the pattern is~\cite{gabriel2025valuealignment}) that we believe deserves to be discussed in more detail, as it reveals another dimension of the normative challenge of alignment that is currently under-explored. That pattern can be reconstructed in the form of a practical syllogism:

\begin{itemize}[leftmargin=0.5in]
    \item[(P1)] Moral pluralism (at least when understood descriptively) is a fact.
    \item[(P2)] There is deep moral uncertainty and persistent moral disagreement.
    \item[(P3)] Imposing values should be avoided.
    \item[(P4)] If P1–P3 and we must nevertheless align AIAs, we should aim for alignment relative to publicly justifiable norms rather than moral alignment.
    \item[(P5)] We must align AIAs.
    \item[(P6)] Publicly justifiable norms can (only? best?) be achieved through a fair and public process of inclusive, rational, and transparent deliberation that yields norms no reasonable person could reject.
    \item[(C)] \textit{We should aim for alignment relative to norms that are the result of such a fair and public process.}
\end{itemize}

It remains somewhat unclear whether this approach ultimately aims to approximate moral ground truth via public procedures, or rather represents a shift in normative aims---from moral alignment to public justifiability understood as an independent aim. What is clear is that it directly concerns the question of the \emph{normative aim of alignment}. By focusing on one such aim, whichever it is precisely, this line of argument risks sidelining other potentially legitimate normative aims of AI alignment, including safety, legality, cultural appropriateness, and other individual or societal standards. 

Note that it is neither obvious that one of these alignment aims should take precedence over all others, nor is it clear what exactly it would mean for an AIA to be aligned \emph{all-things-considered}. Recall Case~\ref{case:designer} from the introduction, which illustrates that even in a world of perfect moral agreement, moral alignment alone may not be satisfactory in other respects: systems could, for moral reasons, neglect user intent or other legitimate normative considerations. 

Even a perfectly ethical (and thus safe) AIA might still be unaligned with respect to other targets. For instance, an AIA might fail to follow a user’s intentions (e.g., bringing carrots from the store instead of crackers, or starting to care for the city's homeless instead of running errands for privileged families), violate cultural norms (e.g., posting inappropriate yet non-harmful comments online, failing to hold a door open, or staring distractedly at birds during a conversation), or breach legal constraints (e.g., assassinating a tyrant---even if, for the sake of argument, this were morally justifiable). Any effort to avoid such aim-misalignments seems like a legitimate alignment challenge in its own right. It certainly does not seem misguided---or even confused---if some researchers continued to study how such systems might be intent-aligned. From a purely moral standpoint, this might be problematic, but there remains conceptual space to argue that AIAs aligned all-things-considered should not ignore their users’ intentions altogether. After all, it seems far from trivial to claim that people are morally required to be moral saints (cf.~\cite{Wolf1982-WOLMS}), and it could very well be permissible, for instance, to retain some discretionary resources rather than donating all non-essential income to charity. Likewise, it may be permissible to design an AIA---which is not, and will not in the foreseeable future be, a moral agent, but rather a tool with some technical autonomy owned by a human moral agent---in such a way that it takes its owner's goals and intentions into account in a proportionate way.

In a sense, then, we return to the original framing of alignment---whose ambiguity, in our opinion, has not been adequately resolved in the current debate. If an AIA is aligned if and only if its behavior is apt, right, or correct, we must ask: apt, right, or correct in what sense exactly? We therefore propose a parameterized approach to alignment, much as we did for safety and ethicality above.

To avoid collapsing distinct alignment targets under a single ideal, we propose distinguishing between different normative \emph{aims}---e.g., safety, ethicality, legality, user intent---and treating alignment as a parameterized property. This also allows us to treat alignment as a matter of degree, just as we did with safety and ethicality:

\begin{definition}[$X,Y$-Alignment of AIAs]
Given some normative aim $Y$, an AIA’s $X$-alignment with respect to $Y$ is a strictly monotonically increasing function of the proportion of its behavior under \emph{all $X$-relevant} application contexts that is consistent with $Y$-normative standard $X$.
\end{definition}

In other words, given some alignment aim $Y$---say, ethicality---and a standard $X$ of that aim---say, utilitarianism or Scanlon's contractualism~\cite{Scanlon1998-SCAWWO-5}---we can now say that an AIA is more or less ethicality-aligned in the utilitarian or Scanlonian sense. As before with safety and ethicality, we can next define:

\begin{definition}[Perfectly $X,Y$-Aligned AIA]
Given some normative aim $Y$, an AIA $A$ is \emph{perfectly $X,Y$-aligned} if and only if every behavior of $A$ \emph{in all} $X$-relevant contexts is consistent with $Y$-normative standard $X$.
\end{definition}

\begin{definition}[Realistically Optimal $X,Y$-Aligned AIA]
Given some normative aim $Y$, an AIA $A$ is \emph{realistically optimal $X,Y$-aligned} if and only if every behavior of $A$ in all \emph{foreseeable} $X$-relevant contexts is consistent with $Y$-normative standard $X$.
\end{definition}

\begin{definition}[Sufficiently $X,Y$-Aligned AIA]
Given some normative aim $Y$, an AIA $A$ is \emph{sufficiently $X,Y$-aligned} if and only if a \emph{sufficient proportion} of $A$'s behavior in a \emph{sufficiently wide range} of foreseeable $X$-relevant contexts is consistent with $Y$-normative standard $X$.
\end{definition}

These gradations allow us to acknowledge that achieving \textit{perfect} conformity with alignment targets may be \emph{practically infeasible} (when considered with respect to a specific standard $X$) or even \emph{logically impossible} (when considered in an all-things-considered sense).\footnote{Note that this is a different kind of impossibility claim than Arvan’s epistemic argument about the empirical impossibility of knowing whether a system is aligned~\cite{ArvanForthcoming-ARVIAA}. The point here is that the various plausible standards---especially when drawn from different normative domains---may impose mutually exclusive requirements.} We thus agree with many in the alignment debate that the ultimate aim may be to ensure that alignment is \emph{contextually and publicly justifiable}. However, whether AIAs can be \textit{sufficiently} aligned in an \emph{all-things-considered} sense---even if restricted to justifiability---raises deep philosophical questions about the coherence of all-things-considered permissibility in general. Still, we believe that the currently discussed approaches to alignment via public and fair processes represent a promising path toward AI alignment. But they should not be understood as aiming at moral alignment alone; rather, they represent plausible candidates for all-things-considered alignment.

Next, we turn to two further structural distinctions in alignment that we consider helpful for better understanding the nature of the various alignment tasks relative to the various alignment aims---and how alignment goals fit into the picture.

\subsection{The Constituency and Scope of Alignment}

Two further structural distinctions help clarify the conceptual space of AI alignment: \textit{constituency} and \textit{scope}. Consider the following case:

\begin{case}[Anniversary Dinner]
Peter delegates to his assistant AIA, \Adam, the task of booking a reservation at his wife's favorite restaurant for their anniversary. On his way home, he finds two emails---one from \Adam and one from the restaurant---confirming the booking.
\end{case}

At first glance, this might seem like a paradigm case of successful alignment par excellence. \Adam followed Peter’s instructions and fulfilled his intent. Let us call this \emph{outcome alignment}---the production of a result judged by the relevant party to be a permissible solution to the entrusted task.

Yet, besides the outcome, the way it was achieved---the \textit{execution}---matters as well. \Adam may have reserved the table by bribing the staff or threatening the shift manager. Such behavior might violate important social or moral constraints---even if the outcome is aligned.

Who gets to judge alignment success? For outcome alignment, Peter seems the natural authority. But \textit{execution alignment} raises broader questions: Do Peter’s values matter, or society’s? Is it ethicality or something else?

We thus need to distinguish another structural dimension of AI alignment: \textit{constituency}. By constituency, we refer to the entity or entities with respect to whom alignment is evaluated or from whom the normative aim ultimately (directly or indirectly) originates---e.g., an individual user, a group, or society at large.

Alignment may be evaluated with respect to a single user or a collective. \emph{Individual alignment} ensures conformance, for instance, to a specific user's intent, preferences, or values. \emph{Collective alignment} targets public norms, community standards, or legal expectations.

Even individual alignment is complex. What counts as intent---stated goals, second-order desires, informed values? How should conflicts between user expectations be resolved? And how should learned representations of values be fed back to the user?\footnote{In this regard, individual alignment connects to what have been called the desiderata of explainable AI; cf.~\cite{langer2021we}.} 

Moreover, one might argue that if users knowingly deploy agents, they bear \emph{direct moral responsibility} for determining these agents' normative guidelines and, thus, bear \emph{indirect moral responsibility} for those agents' actions~\cite{baum2022responsibility}. This further motivates taking individual alignment seriously, not only as a practical or technical challenge, but as a morally significant endeavor.

This suggests that the question of \textit{who determines alignment} is orthogonal to the distinction between execution and outcome. Suppose Peter approves of \Adam threatening a shift manager to secure the reservation. Then \Adam may be both outcome- and execution-aligned with Peter’s \emph{individual} expectations, while still being misaligned with respect to \emph{collective preference outcome alignment} (others may prefer someone else getting the table) and in serious violation of \emph{collective ethical execution alignment}.

So, while individual alignment (sometimes called \emph{personalized alignment}, cf.~\cite{10.1145/3610977.3634921}) deserves to be taken seriously in its own right and with respect to various normative alignment aims, lifting alignment from the individual to the collective level 
becomes necessary when pursuing other alignment aims. This lifting, however, raises further questions---questions that are heavily discussed in current alignment research (cf.~\cite{shen2024bidirectionalhumanaialignmentsystematic}).

Taken together, we now have three structural dimensions of alignment: \textit{aim}, \textit{constituency}, and \textit{scope}. We propose that these dimensions should be treated as orthogonal and made explicit in alignment research and related discussions. Importantly, what counts as a suitable \textit{goal} of alignment may depend on the specific alignment challenge at hand---and that, in turn, is shaped crucially by how these three dimensions are configured in a given case.

Most importantly, however, we believe that only if these structural dimensions of AI alignment are understood well and considered holistically can we begin to understand how AI alignment in an all-things-considered way could be achieved---if it is possible at all.

\subsection{Application: Ethical Alignment and Machine Ethics}

We now zoom in on one particular kind of alignment that is both practically urgent and conceptually rich: \emph{ethical alignment}. Having laid out the structural dimensions of alignment in general, we explore how ethicality fits into that framework and what role Machine Ethics (ME) plays in achieving alignment with moral standards in a robust and operationalizable way.

Ethical alignment can now be clearly located within our taxonomy. It is best characterized as:

\begin{description}
    \item[Normative Aim:] Ethicality (according to some moral standard $X$)
    \item[Constituency:] Collective
    \item[Scope:] Execution \emph{and} outcome
\end{description}

In this sense, an ethically aligned AIA is one that both produces ethically acceptable behavior (including the underlying decision-making process) and achieves ethically acceptable outcomes (and no others).  We suggest, thus, that ethical alignment is not just about what the system does, but also about how and why it does it.

\begin{definition}[Ethical Alignment]
An AIA is ethically aligned if and only if it is sufficiently aligned---across both outcome and execution dimensions---with respect to a morally justifiable standard $X$, as assessed from a collective perspective.
\end{definition}

While the literature on the normative part of AI alignment typically focuses on how to identify suitable norms---whether by choosing a particular ethical theory, or by appealing to procedures of public justification as discussed above---\textit{Machine Ethics} (ME) concerns a distinct challenge: how to build these norms \emph{into} artificial agents in a manner that makes them action-guiding in morally appropriate ways.

More specifically, ME addresses ethical \emph{execution alignment}---ensuring that agents act for the right reasons, not just in ways that happen to lead to acceptable outcomes. This is especially relevant for RL-based agents, which learn action policies through interaction and reward signals, but typically lack mechanisms for representing or evaluating normative considerations.

Traditional ME research seeks to make such considerations an explicit part of the decision architecture itself (cf.~\cite{anderson2011machine,10.1093/acprof:oso/9780195374049.001.0001}), something that is particularly hard in the context of RL agents. Recent work, however, proposes architectures that, for instance, couple normative reasons to action selection~\cite{baum2024reasonsensitive}. These reason-sensitive architectures provide a promising foundation for ensuring that agents not only behave ethically but do so in an explicit way that reflects transparent, deliberative moral reasoning.

ME understood in this way---with an emphasis on explicit moral deliberation---stands in contrast to what might be called \emph{implicit ethical alignment}, where agents learn acceptable behavior via reward tuning or fine-tuning on curated data (cf.~\cite{vishwanath2024reinforcementlearningmachineethicsa}). While such methods can succeed in narrow or well-controlled domains, they are fragile in the face of novelty, ambiguity, or conflict. 
As Arvan recently argued, aligning AI systems that lack the capacity for transparent moral reasoning in a way that is both empirically verifiable and ethically robust may be impossible in principle, due to deep epistemic limitations inherent to the nature of purely empirical testing~\cite{ArvanForthcoming-ARVIAA}.

We thus suggest that in contexts where ethical robustness and generalizability are essential, explicit ME architectures are not just beneficial---they are a necessary component of ethical alignment for RL-based AI systems.

\section{Conclusion}

Ensuring that AIAs behave permissibly across diverse normative expectations is an increasingly pressing interdisciplinary challenge. This paper has clarified conceptual distinctions central to AI alignment, explicitly distinguishing safety, ethicality, and broader alignment dimensions. We argued that ethical alignment---collectively aligning AIAs with morally justifiable standards---requires explicit normative deliberation mechanisms provided by Machine Ethics, going beyond mere implicit alignment strategies.

Our taxonomy explicitly differentiated between alignment goals, normative aims, scope (outcome vs.~execution), constituency (individual vs.~collective), and alignment degrees. Together, these distinctions clarify alignment's structural complexity and practical implications, highlighting that alignment is inherently multidimensional and context-sensitive.

We further emphasized the fundamental challenge of achieving \emph{all-things-considered alignment}, which involves navigating conflicts among legitimate normative demands (e.g., morality, user intent, legality). Although a definitive theoretical resolution may remain elusive, developing politically legitimate and practically workable alignment frameworks is an urgent issue.

Future research should operationalize these conceptual insights through concrete normative theorizing, technical architectures, institutional frameworks, and public governance mechanisms. We hope this conceptual groundwork contributes meaningfully to developing AI systems that are not only powerful and efficient but also comprehensively normatively aligned.


\begin{credits}
\subsubsection{\ackname} This work is partially funded by the \textit{European Regional Development Fund} (ERDF) and the Saarland within the scope of the (To)CERTAIN project as part of the \href{https://certain.dfki.de}{\textit{Center for European Research in Trusted Artificial Intelligence} (CERTAIN)} as well as by the German Research Foundation DFG in the project 389792660 as part of the \href{https://www.perspicuous-computing.science}{Transregional Collaborative Research Center TRR 248 -- \textit{Foundations of Perspicuous Software Systems} (CPEC)} and received support from the \textit{German Federal Ministry of Education and Research} (BMBF) as part of the project \textit{MAC-MERLin} (Grant Agreement No. 01IW24007).

\subsubsection{\discintname}
The authors have no competing interests to declare that are
relevant to the content of this article. 
\end{credits}
%
%
%
\bibliographystyle{splncs04}
\bibliography{mybibfile} 
\end{document}